\documentclass{article}
\usepackage{amsmath,graphicx,mlspconf}
\usepackage{xcolor}
\usepackage{booktabs}
\usepackage{microtype}

\copyrightnotice{U.S.\ Government work not protected by U.S.\ copyright}

\copyrightnotice{979-8-3503-2411-2/25/\$31.00 {\copyright}2025 Crown}

\copyrightnotice{979-8-3503-2411-2/25/\$31.00 {\copyright}2025 European Union}

\copyrightnotice{979-8-3503-2411-2/25/\$31.00 {\copyright}2025 IEEE}

\title{Bridging Discrete and Continuous: A Multimodal Strategy \\for Complex Emotion Detection}
\name{Jiehui Jia\qquad Huan Zhang \qquad  Jinhua Liang\thanks{Anonymous.}}
\address{\textit{Queen Mary University of London,} Centre for Digital Music, London, United Kingdom}

\begin{document}
\ninept

\maketitle

\begin{abstract}
In the domain of human-computer interaction, accurately recognizing and interpreting human emotions is crucial yet challenging due to the complexity and subtlety of emotional expressions. This study explores the potential for detecting a rich and flexible range of emotions through a multimodal approach which integrates facial expressions, voice tones, and transcript from video clips. We propose a novel framework that maps variety of emotions in a three-dimensional Valence-Arousal-Dominance (VAD) space, which could reflect the fluctuations and positivity/negativity of emotions to enable a more variety and comprehensive representation of emotional states. We employed K-means clustering to transit emotions from traditional discrete categorization to a continuous labeling system and built a classifier for emotion recognition upon this system.  The effectiveness of the proposed model is evaluated using the MER2024 dataset, which contains culturally consistent video clips from Chinese movies and TV series, annotated with both discrete and open-vocabulary emotion labels. Our experiment successfully achieved the transformation between discrete and continuous models, and the proposed model generated a more diverse and comprehensive set of emotion vocabulary while maintaining strong accuracy.
\end{abstract}


\section{Introduction} \label{sec:introduction}
Human emotions are complex and described through diverse vocabularies across languages, reflecting our thoughts, feelings, and reactions via facial expressions, body language, voice tone, and speech.~\cite{pally1998emotional}. Accurate comprehension and response to human emotions by machines can significantly benefit areas such as marketing, mental health monitoring, multimedia generation, and human-computer interaction~\cite{esposito2015needs, zhang2025aestheticshumanpreferencescomparative, thieme2020machine, mariani2022ai, app14156543, zhang2025renderbox}. Thus, developing systems capable of precisely identifying varieties of human emotions is essential.

However, the challenge in emotion detection lies in the subjective nature of emotions. It is hard to set a clear boundary to categorize emotions, so as to choose a `basic' emotion group~\cite{prinz2004emotions}. Moreover, emotion datasets vary in their annotation schemes (e.g., differing discrete labels) and domains, hindering direct comparisons across previous works. These variations restrict prior research to specific data sources, limiting their generalizability to real-world applications~\cite{oberlander2018analysis}. 

\begin{figure}[htb]
    \includegraphics[width=\linewidth]{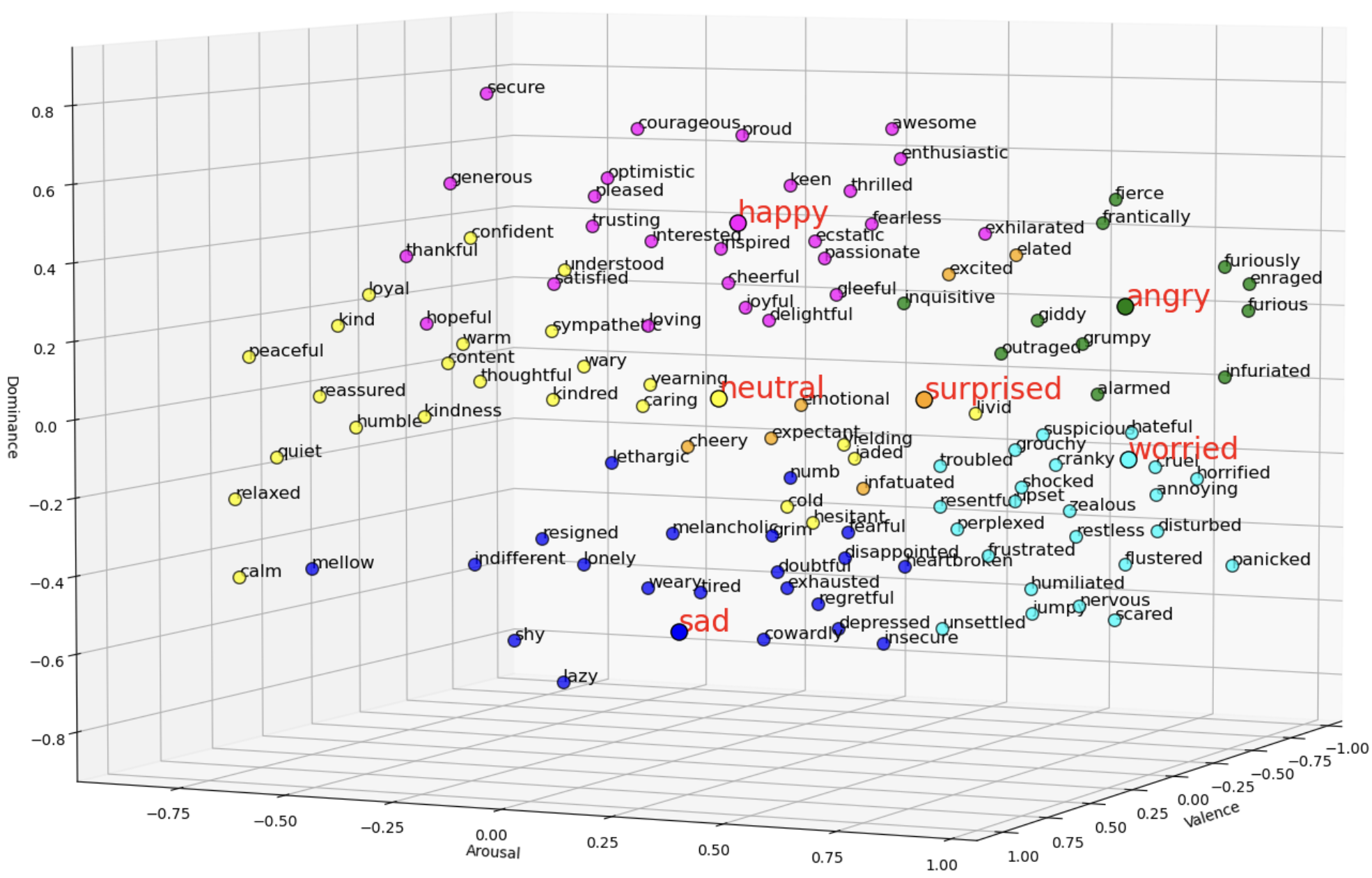}
    \caption{Emotion Vocabularies in 3D VAD Space}
    \label{fig:Emotion Words in 3D VAD Space}
\vspace{-0.5cm}
\end{figure}

Recognizing the challenges, this paper proposes a multimodal framework that transforms different discrete emotion labels into one continuous emotion label framework. We propose to use a fixed emotion Valence, Arousal, and Dominance rating scale to standardize the scoring of a variety of emotion labels in a multidimensional space. By using K-means clustering, we build a classifier which could transit discrete emotion with continuous emotions which suits both close-set and open set emotion recognition. In our case, we grouped the emotion labels into six clusters based on the six basic emotion labels which been annotated for the dataset we chosen. We built a multimodal model that integrates facial expressions, voice tones, and transcript from video clips to predict the Valence-Arousal-Dominance (VAD) scores of the emotions. Finally the VAD score was mapped back to the original emotion labels for evaluation. We also generated open-vocabulary emotional responses from discrete emotion inputs to explore the possibility of a more dynamic and adaptable emotion recognition system.

The experimental results demonstrate that the framework successfully achieved the transformation between discrete and continuous models. By comparing their evaluation metrics, we conclude that the proposed framework performs the transformation with a comparable level of accuracy. We also compared the similarity score between the generated open emotion vocabulary set and the original dataset's open set, achieving a high score that indicates significant semantic overlap. This shows that our model can reliably produce a nuanced and flexible emotion vocabulary, which could be effectively applied to tasks requiring more detailed emotion recognition and analysis. The contribution of this paper is summarized as follows:

\begin{itemize}
    \item We introduce a multimodal system that aligns with human perception by mapping emotions into a continuous hidden space. The proposed framework outperforms existing classifier in the close-set dataset.
    \item We incorporate VAD rating system with an emotion classifier, which learns a more nuance representation of emotion states than learning from discrete categories.
    \item We benchmark the open-set emotion classification task by applying the wav2vec model. Experiments demonstrate a high correlation between the ground truth and our proposed model.
\end{itemize}

\section{Related work} \label{sec:related_work}
Compared to regular classification tasks~\cite{10446908,liang2024mind,zhang2022adaptive}, multimodal emotion detection methods primarily differ in two key areas: the approaches to categorizing and measuring emotions, and the techniques used to extract and integrate features from various modalities.

\vspace{-0.15cm}\subsection{The measurement of emotion} \label{subsec:emotion_measure}

There are two main approaches to measuring emotions: the discrete method and the multidimensional approach~\cite{KHARE2024102019}. The discrete method classifies emotions into a few fundamental categories believed to be universal. Proponents argue that combinations of these basic emotions can explain complex emotional states. For instance, fear, surprise, happiness, disgust, anger, and sadness has been identified as the six basic emotions~\cite{ekman1971constants}.

Building upon these basic feelings, a thorough emotional model known as Plutchik's wheel of emotions has been developed~\cite{plutchik2001nature}, which consists of eight main emotions. It is believed that different intensities of these main emotions might mix to create additional related feelings. Studies have been conducted on dataset with 15 Compound emotion mixed with two basic emotions and 7 Basic facial emotions as a way for compound emotion detection~\cite{compoundemotion}. In some situations, the discrete method works well enough with a simplified emotional assessment~\cite{mitchell2023_deep,9360728}. For example, when it comes to driving systems in vehicles, determining a driver's stress level can be accomplished by concentrating just on the most fundamentally feelings, like anger and happiness~\cite{9360728}. 

\begin{figure*}[htb]
\centering
\includegraphics[width=0.9\linewidth]{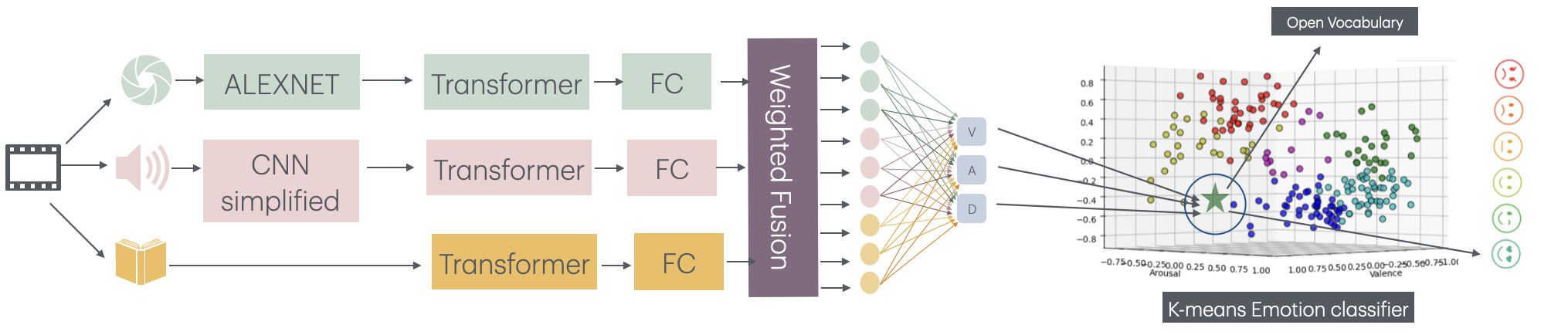}
\caption{VAD Model Pipline}
\label{fig:VAD Model Pipline}
\vspace{-0.2cm}
\end{figure*}

The multidimensional approach, on the other hand, uses continuous dimensions—valence, arousal, and sometimes dominance—to map emotions. Valence measures positivity or negativity, arousal indicates the level of activity, and dominance reflects control over the emotional state~\cite{ps2017emotion}. This creates a nuanced two- or three-dimensional emotional space, often referred to as the circumplex model of affect~\cite{russell1980circumplex}.

\vspace{-0.15cm}\subsection{Multimodal Emotion Detection} \label{subsec:multimodal_emotion_detection}
Multimodal methods have been widely chosen in research because they can extract features from various modalities~\cite{kazemzadeh2013natural,lian2023survey,liang23c_interspeech,liang2023acousticprompttuningempowering}. While basic emotions like happy or angry are readily expressed through body language or facial expressions, complex emotions such as jealousy, pride, or hope require additional context and language for accurate interpretation~\cite{kazemzadeh2013natural}. Multimodal methods are thus preferable to unimodal ones for recognizing such emotions. This approach allows researchers to analyze data from different sources—such as visual, audio, and textual inputs—in ways that enhance the accuracy and robustness of emotion detection systems. By utilizing these different streams of information, multimodal methods capture a more comprehensive understanding of emotional states, accommodating variations and subtleties that might be missed by unimodal approaches~\cite{lian2023survey}.

Visual, audio, and textual features are captured separately using different methods. Visual features primarily focus on facial expressions, which can be seen as facial muscle movements.
The Facial Action Coding System (FACS)~\cite{ekman1978facial}  has been developed to manually code facial expressions using action units (AUs) based on specific muscle movements. Inspired by their work, many researchers have utilized image and video processing to analyze and categorize facial expressions by tracking features and measuring movement, applying these methods to the ``basic expressions" identified in multimodal emotion studies~\cite{zhang2016multimodal,keltner1798understanding,metri2011facial,thushara2016multimodal}. Similar to other audio domains~\cite{DING2024121902, zhang2025renderbox}, emotion-related features consist of two main categories: linguistic information, which pertains to the content of speech, and paralinguistic information, which captures emotions through the tone and manner of delivery~\cite{el2011survey}. In addition to commonly used models like CNN, RNN, and LSTM~\cite{lim2016speech,satt2017efficient}, other approaches also make promising prediction. For instance, the use of features such as mel-frequency cepstral coefficient (MFCC), perceptual linear prediction coefficient (PLPC), and perceptual minimum variance distortionless response (PMVDR) coefficient~\cite{daneshfar2020speech} . Textual emotions are expressed through speech and can be abstract by transcripts. Traditional techniques such as the bag-of-words (BoW) model~\cite{schmitt2016border,spyrou2017extracting} have been commonly used. Then word embedding methods like tf-idf, word2vec~\cite{cahyani2021performance} and GloVe~\cite{gupta2021decoding} were based mainly on syntactic contexts. For handling larger text, BERT (Bidirectional Encoder Representations from Transformers) uses transformer-based encoders that assess both preceding and subsequent contexts to better predict words in sentences~\cite{devlin2018bert}.

\section{Methodology} \label{sec:method}
The project aims to bridge discrete and continuous emotion systems, using a multimodal model to detect emotions and produce more nuanced emotion labels beyond the basic six. This approach allows for working with diverse datasets and improves the variety and depth of emotion detection. We first transformed discrete emotion labels into continuous emotion labels, then applied and refined several models for the task. The first model, Multimodal End-to-End Sparse Model (ME2E)~\cite{dai-etal-2021-multimodal}, is a multimodal emotion detection model that uses discrete emotion labels as input and outputs corresponding discrete emotions, serving as our baseline. The second, ME2E Lite model, a refined version of ME2E to better align with our dataset for improved performance. The third, Proposed VAD, builds on the ME2E Lite pipeline but uses 3D VAD score as input, outputting continuous VAD scores that can be mapped to discrete emotions while also generating a broader range of nuanced emotion vocabularies.

\vspace{-0.2cm}\subsection{The K-means clustering classifier - Transformation between discrete emotion labels and continuous emotion labels} \label{subsec:model_transformation}
To transition from discrete to multidimensional emotion methods, we developed a classifier that enables the transformation between discrete and continuous emotion labels using the NRC-VAD lexicon~\cite{mohammad2018obtaining}. We extracted 195 emotion vocabularies with assigned VAD scores from the 20,000-vocabularies NRC-VAD lexicon, using the polar term based on a -1 to 1 scale, as taking the absolute values of these scores is likely to yield better features than those derived from a 0 to 1 scale. The six basic emotions selected—happy, sad, worried, surprised, angry, and neutral—align with the categories used in our dataset. As VAD scores are available for all selected basic emotions except neutral, we assigned a VAD score of (0,0,0) to neutral, reflecting a lack of significant emotional fluctuations typically associated with this state.

At this stage, each basic emotion can be transformed into a 3D VAD score, but to enable the inversion from the continuous emotions to discrete labels, we employed a K-means clustering classifier. By setting k to 6, the classifier is configured to convert continuous VAD scores back into six basic emotion categories. 

To build the emotion classifier, we first mapped the 195 extracted emotion vocabularies, along with their associated VAD scores, into a 3D emotion space. The six basic emotions—happy, sad, worried, surprised, angry, and neutral—were set as the initial cluster centers. We then set the number of clusters to six, corresponding to the number of basic emotions.

The underlying assumption is that people experiencing similar emotions will exhibit similar VAD scores, meaning their emotional reactions should cluster closely in the VAD space. By applying K-means clustering, we grouped these similar emotions into clusters. Each cluster represents a collection of VAD scores that align with one of the basic emotions. After clustering, the centroid of each cluster corresponds to a discrete emotion, allowing us to map continuous VAD scores back to a specific emotion category.

This approach effectively allows us to bridge continuous and discrete emotional labels, with the results of the clustering visualized in Figure~\ref{fig:Emotion Words in 3D VAD Space} and Table~\ref{tab:emotion_cluster}, demonstrating how similar emotions are grouped together based on their VAD scores.

\vspace{-0.2cm}
\title{Emotion Clustering}
\maketitle

\begin{table}[ht]
\caption{Emotion Clustering}
\label{tab:emotion_cluster}
\centering
\begin{tabular}{c c c c c}
\toprule
Cluster & Examples in Cluster\\ [0.5ex] 
\midrule
Happy & delighted, inspired, glad, humorous, cheerful  \\
Sad & fatigued, mournful, vulnerable, doubtful, regretful  \\
Worried & guilty, offended, wounded, annoyed, frightened  \\
Neutral & kind, warm, thoughtful, sympathetic, humble  \\
Surprised &  emotional, elated, expectant, curious, impressed  \\
Angry & grumpy, vengeful, moody, offended, frantic
\\ [1ex]
\bottomrule
\end{tabular}
\vspace{-0.4cm}
\end{table}

\vspace{-0.15cm}\subsection{ME2E - Baseline Model} \label{subsec
} The ME2E model processes data from video, audio, and text modalities. Facial features are extracted from video frames using the MTCNN model from FaceNet~\cite{schroff2015facenet}, while both these facial features and audio spectrograms are then processed through convolutional layers. These features are further analyzed using a Transformer~\cite{vaswani2017attention} to capture temporal and contextual nuances. Text data is processed through ALBERT~\cite{lan2020albertlitebertselfsupervised}, which optimized for short sentences and requiring fewer parameters than BERT. Finally, features from each modality are individually processed and merged via a weighted fusion mechanism, producing a final emotion prediction across six possible outcomes: happy, angry, sad, neutral, worried, or surprised.

\vspace{-0.15cm}\subsection{ME2E Lite - The refined Model} 
Building on the work of~\cite{dai-etal-2021-multimodal}, we revised the architecture in our ME2E Lite model by replacing the original 11-layer CNN+VGG video path with AlexNet~\cite{krizhevsky2012imagenet}, and simplified the audio path’s CNN by removing one VGG layer, effectively halving the processed parameters to better suit our smaller dataset.

\vspace{-0.15cm}\subsection{Proposed VAD Model – The model bridges the gap between discrete and continuous emotions}

The proposed VAD model shown in Figure~\ref{fig:VAD Model Pipline} uses VAD scores as input, making it compatible with both continuous labels and discrete emotions which been represented by VAD scores using the NRC-VAD lexicon. We employed the same pipeline as ME2E Lite but replaced the softmax function with mean squared error (MSE) as the loss function. The model features a fully connected layer that directly outputs three continuous values corresponding to the VAD dimensions

This model independently processes each of these dimensions by predicting the distribution for valence, arousal, and dominance, then identifying the most likely score for each. These predicted VAD scores are then mapped to the K-means clustering classifier. By determining which emotion cluster each result falls into, we can transform the continuous VAD scores into discrete emotion labels for evaluation. This allows for a comparison of the continuous emotion model’s performance against the discrete emotion model. Alternatively, by selecting the emotion close to the predicted VAD score, we can generate a broader, open set of possible emotions, offering more nuanced emotion representations.

\section{Dataset} \label{subsec:implementation}
We have selected the MER2024 dataset~\cite{lian2024mer} for our study. This dataset consists of raw video clips collected from Chinese movies and TV series, ensuring cultural consistency. The raw video samples were split into smaller video clips, each containing a complete segment from the same character. The dataset has been labeled with six discrete emotions: happy, angry, sad, neutral, worried, and surprised. Additionally, there is a subset of open-vocabulary labels and explanations~\cite{lian2023explainable}, which allows us to evaluate a more dynamic emotion output. Each sample contains audio data with a sampling rate of 44.1 kHz, text transcript and video frames. The video operates at 25 frames per second (FPS), and the frames are sampled every 500 milliseconds which result in capturing about two frames each second.We created a dataloader to split the data into training (70\%), validation (15\%), and test sets (15\%).The details of our data split are shown in Table~\ref{tab:emotion_data}.

\begin{table}[ht]
    \centering
    \caption{Dataset Distribution. C-Avg and W-Avg represent the average of clip duration and word counts, respectively.}
    \label{tab:emotion_data}
    \begin{tabular}{lcccccc}
    \toprule
    Emotion & Total & Train & Val & Test & C-Avg& W-Avg\\ \midrule
    Happy   & 931     & 538   & 120 & 113  & 3.51s             & 12.18    \\
    Angry   & 1100    & 683   & 141 & 116  & 3.90s            & 15.49    \\
    Sad     & 564     & 282   & 42  & 80   & 5.66s             & 12.70    \\
    Neutral & 1142    & 679   & 153 & 150  & 3.81s             & 13.88    \\
    Worried & 585     & 282   & 75  & 68   & 4.54s             & 15.23    \\
    Surprised & 183   & 108   & 20  & 25   & 2.92s             & 8.05     \\
    \bottomrule
    \end{tabular}
    \vspace{-0.7cm}
    \end{table}

\section{Experiments setup and Evaluation} \label{sec:experiments}
For all three models, the SGD optimizer was employed. To mitigate initial overfitting, batch normalization and dropout layers were added after each convolution layer, coupled with ReLU activation functions to add non-linearity. The learning rate was dynamically adjusted during training using the CosineAnnealingLR scheduler to help the model steer clear of local minima. The model has been trained on a single NVIDIA Tesla V40 GPU with 48GB of memory, spanning 30 epochs and completing in about 1.2 hours.

\textbf{The ME2E and ME2E Lite models} We used a batch size of 32, achieving the best performance with a learning rate of 0.0001 and a weight decay of 0.005. As it is a single-label, multi-class task for these two models, we employed the Cross-Entropy loss function, given by $L = -\sum_{i=1}^{N} y_i \log(p_i)$, where $L$ is the loss for an example, $N$ is the number of classes, $y_i$ is the binary indicator for correct class classification, and $p_i$ is the predicted probability for class $i$. We applied Softmax  activation which ensures a probability distribution over classes for classification. Evaluation metrics include precision, recall, F1 score, and Accuracy.

\textbf{The Proposed VAD model} We used a batch size of 8, reaching optimal performance with a learning rate of 0.0005 and a weight decay of 0.0001. Since it is a continuous task, we used Mean Squared Error (MSE) as the loss function and evaluation metric.We also evaluate using L2 distance, Mean Absolute Error (MAE), and Pearson Correlation Coefficient (PCC) to assess the linear relationship between predictions and actual values, where 
\begin{equation*}
    \text{PCC} = \frac{\sum_{i=1}^N (y_i - \mu_y)(\hat{y}_i - \mu_{\hat{y}})}{\sqrt{\sum_{i=1}^M (y_i - \mu_y)^2} \cdot \sqrt{\sum_{i=1}^M (\hat{y}_i - \mu_{\hat{y}})^2}}.
\end{equation*}

\section{Results Analysis} \label{sec:experiments}

We have trained the baseline model, ME2E Lite model and the proposed VAD model on the MER2024 dataset and evaluated their performance using the metrics described above. 

\vspace{-0.25cm}\subsection{Model performance on continuous emotion detection} \label{subsec:performance_on_continuous}
Table~\ref{tab:Continuous_Result} illustrates the performance of proposed VAD model which deals with the continuous VAD emotion labels. The L2 distance of 0.64 suggests that the model's predictions are relatively close to the actual values, while the MSE of 0.19 and MAE of 0.36 indicate that the model's predictions are generally accurate. The PCC of 0.47 further confirms that the model has learned to predict the emotional values and has achieved a reasonable level of correlation between the predicted and actual values.

\vspace{-0.2cm}
\begin{table}[ht]
    \centering
    \caption{Result on Continuous Emotion Detection}
    \label{tab:Continuous_Result}
    \begin{tabular}{lcccccc}
    \toprule
    Model & L2 distance & MSE & MAE & PCC \\ \midrule
    Raw VAD Model  &  0.96  &  0.38  &  0.48  & -0.01 \\
    Proposed VAD model   &  0.64  &  0.19  &  0.36  & 0.47 \\
    \bottomrule
    \end{tabular}
    \end{table}
\vspace{-0.5cm}

\vspace{-0.15cm}\subsection{Model comparison on discrete emotion detection} \label{subsec:performance_on_discrete}
To assess the performance on discrete emotion detection, we transformed the continuous VAD outputs from the proposed VAD model back into discrete emotion labels. We then evaluated these results alongside the ME2E and ME2E Lite models using F1 score, precision, and recall as performance metrics.

Table~\ref{tab:Discrete_Result} shows that the the proposed VAD model stands out with higher precision (0.49) and recall (0.45), suggesting it is more effective in accurately identifying and capturing emotions. ME2E Lite model also performs well, with an F1 score of 0.42, and balanced precision and recall at 0.42 and 0.43, respectively. The baseline ME2E model had the lowest performance, with an F1 score of 0.33 and lower precision and recall (both at 0.32).

While both the ME2E Lite and proposed VAD models share the same F1 score of 0.42, the VAD model's higher precision indicates it predicts relevant emotions more accurately, and its higher recall means it identifies a greater proportion of true positive emotions. 

\vspace{-0.2cm}
\begin{table}[ht]
    \centering
    \caption{Result on Discrete Emotion Detection}
    \label{tab:Discrete_Result}
    \begin{tabular}{lcccccc}
    \toprule
    Model & F1 & Precision & Recall \\ \midrule
    ME2E (Baseline)     & 0.33  & 0.32& 0.32 \\
    ME2E Lite    & 0.42   & 0.42  & 0.43 \\
    Proposed VAD model   & 0.42   & 0.49   & 0.45 \\
    \bottomrule
    \end{tabular}
    \end{table}
\vspace{-0.5cm}

\vspace{-0.15cm}\subsection{Open-vocabulary exploration} \label{subsec:open_vocabulary}   
At the final stage of our study, we explored the possibility of generating open-vocabulary emotional responses based on six discrete emotion classes. For this, we utilized a subset of 68 samples from the MER2024 dataset which contains the open-vocabulary outputs. We then applied our proposed VAD model to this subset and mapped the resulting VAD labels into the 3D emotion word space showed in the Introduction. To output the final dataset, we used an L2 distance threshold of 0.25, calculated based on the distance which could output an average of five emotion vocabularies from our 3D Emotion Vocabulary Space.

Our finds 
indicate that our model effectively predicts nuanced emotional states. For instance, in Sample 00000368, the MER2024 dataset listed emotions like 'Alert,' 'Excited,' 'Confused,' and 'Curious,' and our model predicted 'Shocked.' This aligns well as 'Shocked' can encapsulate alertness, excitement, and confusion. In Sample 0002419, the dataset included 'Calm,' 'Relaxed,' and 'Happy,' while our model suggested 'Caring' and 'Curious.' These predictions are compatible, with 'Caring' reflecting a relaxed and content state. Full output list of To further examine the alignment of our open vocabulary, we assess the semantic similarity across various model's output of emotion labels.
\section{Conclusion} \label{sec:conclusion} 
Emotions, inherently complex, can be effectively analyzed through a three-dimensional representation, offering a nuanced approach to categorization adaptable to various needs and contexts. The spatial representation within the VAD framework allows for precise, quantitative analysis and easier conversion between different labeling systems, enhancing our understanding of emotional expressions across multiple modalities.

However, our study has limitations, including suboptimal model performance and a small dataset that may hinder generalization. The reliance on the NRC-VAD lexicon, primarily based on English, introduces potential biases in other linguistic or cultural settings. 

Future work could focus on enhancing model performance through larger, more diverse datasets, advanced modeling techniques, and cross-cultural adaptations to better capture the variability in emotional expressions. Further exploration into multimodal integration and more flexible emotion labeling could also provide deeper insights and improve model robustness. Lastly, applying time series analysis to track emotional shifts over time could offer valuable perspectives on emotional dynamics.

\bibliographystyle{IEEEtran}
\bibliography{main}


\end{document}